\documentclass{ws-procs975x65}

\begin{document}

\title{PHOTON ``MASS'' AND ATOMIC LEVELS IN A SUPERSTRONG MAGNETIC FIELD}

\author{M. I. VYSOTSKY$^*$}

\address{A.I. Alikhanov Institute of Theoretical and Experimental Physics,\\
Moscow 117218 Russia\\
$^*$E-mail: vysotsky@itep.ru}

\begin{abstract}
The structure of atomic levels originating from the lowest Landau
level in a superstrong magnetic field is analyzed. The influence
of the screening of the Coulomb potential on the values of
critical nuclear charge is studied.
\end{abstract}


\bodymatter

\vspace{7mm}

It is a great pleasure for me to present this paper to Sergei
Gaikovich Matinyan on his 80th birthday.

\section{Introduction}

We will discuss the modification of the Coulomb law and atomic spectra in
superstrong magnetic field. The talk is based on papers
\cite{1,2,3}, see also \cite{4}.

\section{\mbox{\boldmath$D=2$} QED}

Let us consider two dimensional QED with massive charged fermions.
The electric potential of the external point-like charge equals:
\begin{equation}
{\bf\Phi}(k) = -\frac{4\pi g}{k^2 + \Pi(k^2)} \; , \label{1}
\end{equation}
where $\Pi(k^2)$ is the one-loop expression for the photon polarization
operator:
\begin{equation}
\Pi(k^2) = 4g^2\left[\frac{1}{\sqrt{t(1+t)}}\ln(\sqrt{1+t} +\sqrt
t) -1\right] \equiv -4g^2 P(t) \;\; ,  \label{2}
\end{equation}
and $t\equiv -k^2/4m^2$, $[g] =$ mass.

In the coordinate representation for $k= (0, k_\parallel)$ we obtain:
\begin{equation}
{\bf\Phi}(z) = 4\pi g \int\limits^\infty_{-\infty} \frac{e^{i
k_\parallel z} dk_\parallel/2\pi}{k_\parallel^2 + 4g^2
P(k_\parallel^2 /4m^2)} \;\; . \label{3}
\end{equation}
With the help of the interpolating formula
\begin{equation}
\overline{P}(t) = \frac{2t}{3+2t} \label{4}
\end{equation}
the accuracy of which is better than 10\% for $0<t<\infty$ we
obtain:
\begin{eqnarray}
{\bf\Phi} & = & 4\pi g\int\limits^{\infty}_{-\infty} \frac{e^{i
k_\parallel z} d k_\parallel/2\pi}{k_\parallel^2 +
4g^2(k_\parallel^2/2m^2)/(3+k_\parallel^2/2m^2)} = \nonumber
\\
&=& \frac{4\pi g}{1+ 2g^2/3m^2}\left[-\frac{1}{2}|z| +
\frac{g^2/3m^2}{\sqrt{6m^2 + 4g^2}} {\rm exp}(-\sqrt{6m^2
+4g^2}|z|)\right] \;\; .  \label{5}
\end{eqnarray}

In the case of heavy fermions ($m\gg g$) the potential is given by
the tree level expression; the corrections are suppressed as
$g^2/m^2$.

In the case of light fermions ($m \ll g$):
\begin{equation}
{\bf\Phi}(z)\left|
\begin{array}{l}
~~  \\
m \ll g
\end{array}
\right. = \left\{
\begin{array}{lcl}
\pi e^{-2g|z|} & , & z \ll \frac{1}{g} \ln\left(\frac{g}{m}\right) \\
-2\pi g\left(\frac{3m^2}{2g^2}\right)|z| & , & z \gg \frac{1}{g}
\ln\left(\frac{g}{m}\right) \;\; .
\end{array}
\right. \label{6}
\end{equation}
$m=0$ corresponds to the Schwinger model; photon gets a mass due
to a photon polarization operator with massless fermions.

\section{Electric potential of the point-like charge in
\mbox{\boldmath$D=4$} in superstrong \mbox{\boldmath$B$}}

We need an expression for the polarization operator in the external
magnetic field $B$. It simplifies greatly for $B\gg B_0 \equiv
m_e^2/e$, where $m_e$ is the electron mass and we use Gauss units,
$e^2 = \alpha = 1/137...$. The following results were obtained in
\cite{2}:
\begin{equation}
{\bf\Phi}(k) =\frac{4\pi e}{k_\parallel^2 + k_\bot^2 + \frac{2 e^3
B}{\pi} {\rm exp}\left(-\frac{k_\bot^2}{2eB}\right)
P\left(\frac{k_\parallel^2}{4m^2}\right)} \;\; , \label{7}
\end{equation}
\begin{eqnarray}
{\bf\Phi}(z) & = & 4\pi e \int\frac{e^{ik_\parallel z} d
k_\parallel d^2 k_\bot/(2\pi)^3}{k_\parallel^2 + k_\bot^2 +
\frac{2 e^3B}{\pi} {\rm
exp}(-k_\bot^2/(2eB))(k_\parallel^2/2m_e^2)/(3+k_\parallel^2/2m_e^2)}
= \nonumber \\
& = & \frac{e}{|z|}\left[ 1-e^{-\sqrt{6m_e^2}|z|} +
e^{-\sqrt{(2/\pi) e^3 B + 6m_e^2}|z|}\right] \;\; . \label{8}
\end{eqnarray}

For $B \ll 3\pi m^2/e^3$ the potential is Coulomb up to small
corrections:
\begin{equation}
{\bf\Phi}(z)\left| \begin{array}{l}
~~  \\
e^3 B \ll m_e^2
\end{array}
\right. = \frac{e}{|z|}\left[ 1+ O\left(\frac{e^3
B}{m_e^2}\right)\right] \;\; , \label{9}
\end{equation}
analogously to $D=2$ case with substitution $e^3B \rightarrow
g^2$.

For $B\gg 3\pi m_e^2/e^3$ we obtain:
\begin{equation}
{\bf\Phi}(z) = \left\{
\begin{array}{lll}
\frac{e}{|z|} e^{(-\sqrt{(2/\pi) e^3 B}|z|)} \; , \;\;
\frac{1}{\sqrt{(2/\pi) e^3 B}}\ln\left(\sqrt{\frac{e^3 B}{3\pi
m_e^2}}\right)>|z|>\frac{1}{\sqrt{e B}}\\
\frac{e}{|z|}(1- e^{(-\sqrt{6m_e^2}|z|)}) \; , \;\;  \frac{1}{m_e}
> |z|
> \frac{1}{\sqrt{(2/\pi)e^3 B}}\ln\left(\sqrt{\frac{e^3 B}{3\pi
m_e^2}}\right) \\
\frac{e}{|z|}\;\; , \;\;\;\;\;\;\;\;\;\;\;\;\;\;\;\;\;\;\;\;\;\;\;
  |z| > \frac{1}{m_e}
\end{array}
\right. \;\; , \label{10}
\end{equation}

\begin{equation}
  V(z) = - e{\bf\Phi}(z) \;\; .
\label{11}
\end{equation}

The close relation of the radiative corrections at $B>>B_0$
in $D=4$
to the radiative corrections in $D=2$ QED allows to prove
that just like in $D=2$ case
higher loops are not essential (see, for example, \cite{VB}).

\section{Hydrogen atom in the magnetic field}

For $B > B_0 = m_e^2/e$ the spectrum of Dirac equation consists of
ultrarelativistic electrons with only one exception: the electrons
from the lowest Landau level (LLL, $n=0, \;\; \sigma_z = -1$) are
nonrelativistic. So we will find the spectrum of electrons from
LLL in the screened Coulomb field of the proton.

The wave function of electron from LLL is:
\begin{equation}
R_{0m}(\rho) = \left[\pi(2a_H^2)^{1+|m|} (|m|!)\right]^{-1/2}
\rho^{|m|}e^{(im\varphi - \rho^2/(4a_H^2))} \;\; , \label{12}
\end{equation}

where $m=0, -1, -2$ is the projection of the electron orbital momentum
on the direction of the magnetic field.

For $a_H \equiv 1/\sqrt{eB} \ll a_B = 1/(m_e e^2)$ the adiabatic
approximation is applicable and the wave function looks like:
\begin{equation}
\Psi_{n 0 m -1} = R_{0m}(\rho)\chi_n(z) \;\; , \label{13}
\end{equation}
where $\chi_n(z)$ satisfy the one-dimensional Schr\"{o}dinger
equation:
\begin{equation}
\left[-\frac{1}{2m_e} \frac{d^2}{d z^2} + U_{eff}(z)\right]
\chi_n(z) = E_n \chi_n(z) \;\; . \label{14}
\end{equation}
Since screening occurs at very short distances it is not important
for odd states, for which the effective potential is:
\begin{equation}
U_{eff} (z) = -e^2\int\frac{|R_{0m}(\rho)|^2}{\sqrt{\rho^2 +
z^2}}d^2 \rho \;\; , \label{15}
\end{equation}
It equals the Coulomb potential for $|z|\gg a_H$ and is regular at
$z=0$.

Thus the energies of the odd states are:
\begin{equation}
E_{\rm odd} = -\frac{m_e e^4}{2n^2} + O\left(\frac{m_e^2
e^3}{B}\right) \; , \;\; n = 1,2, ... \;\; , \label{16}
\end{equation}
and for the superstrong magnetic fields $B > m_e^2/e^3$ they
coincide with the Balmer series with high accuracy.

For even states the effective potential looks like:
\begin{equation}
\tilde U_{eff} (z) = -e^2\int  \frac{|R_{0m}(\vec{\rho})|^2}
{\sqrt{\rho^2 +z^2}} d^2\rho \left[1-e^{-\sqrt{6m_e^2}\;z} +
e^{-\sqrt{(2/\pi)e^3 B + 6m_e^2}\;z}\right]
 \;\; . \label{17}
\end{equation}

\begin{center}
\begin{figure}
\includegraphics[width=.8\textwidth]{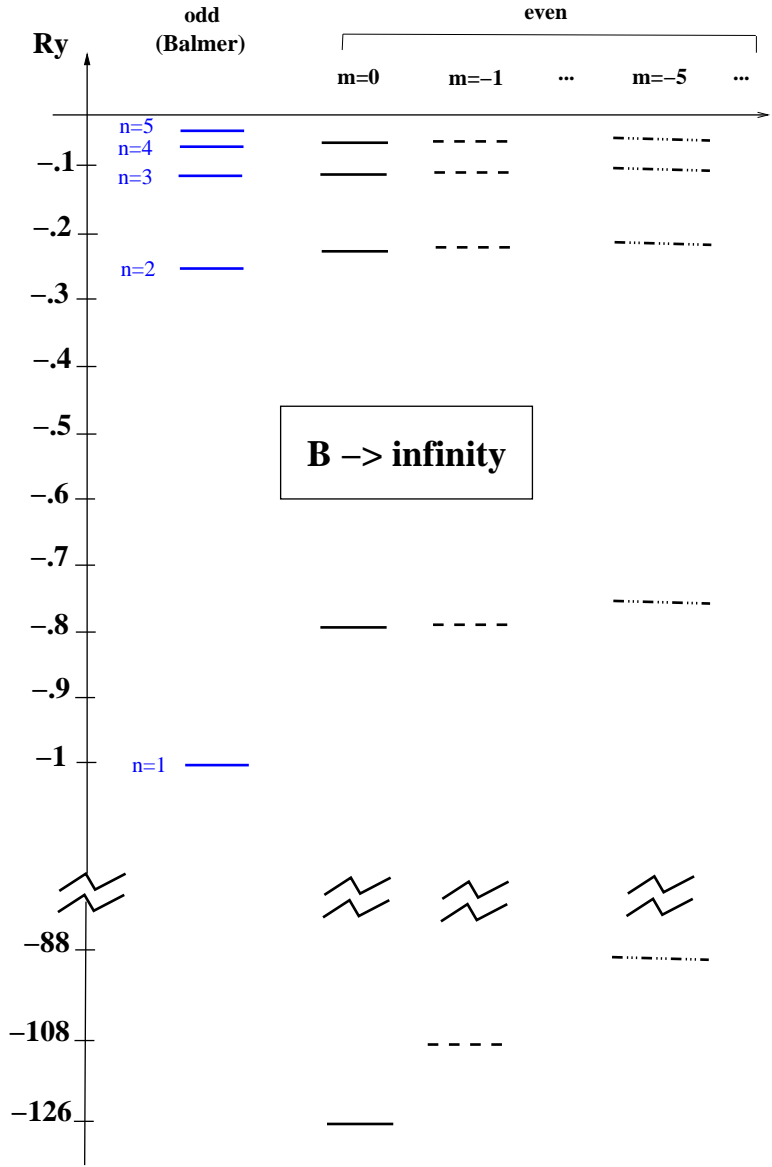}
\caption{Spectrum of the hydrogen atom in the limit of the
infinite magnetic field. Energies are given in Rydberg units, $Ry
\equiv 13.6 \; eV$.}
\end{figure}
\end{center}

Integrating the Schr\"{o}dinger equation with the effective
potential from $x=0$ till $x=z$, where $a_H \ll z \ll a_B$, and
equating the obtained expression for $\chi^\prime(z)$ to the
logarithmic derivative of Whittaker function -- the solution of
the Schr\"{o}dinger equation with Coulomb potential -- we obtain
the following equation for the energies of even states:
\begin{equation}
\ln\left(\frac{H}{1+\displaystyle\frac{e^6}{3\pi}H}\right) =
\lambda + 2\ln\lambda + 2\psi\left(1-\frac{1}{\lambda}\right) +
\ln 2 + 4\gamma + \psi(1+|m|) \;\; , \label{18}
\end{equation}
where $H\equiv B/(m_e^2 e^3)$, $\psi(x)$ is the logarithmic derivative
of the gamma-function and
\begin{equation}
E=-(m_e e^4/2)\lambda^2 \;\; . \label{19}
\end{equation}
The spectrum of the hydrogen atom in the limit $B\gg m_e^2/e^3$ is
shown in Fig. 1.

\section{Screening versus critical nucleus charge}

Hydrogen-like ion becomes critical at $Z\approx 170$: the ground level
reaches lower continuum, $\varepsilon_0 = -m_e$, and two $e^+ e^-$
pairs are produced from vacuum. Electrons with the opposite spins
occupy the ground level, while positrons are emitted to infinity
\cite{ZP}. According to \cite{ORS} in the strong magnetic field
$Z_{\rm cr}$ diminishes: it equals approximately 90 at $B=100
B_0$; at $B= 3\cdot 10^4 B_0$ it equals approximately 40.
Screening of the Coulomb potential by the magnetic field acts in the opposite
direction and with account of it larger magnetic fields are needed
for a nucleus to become critical.

The bispinor which describes an electron on LLL is:
$$
\psi_{e}=\left(\varphi_{e}\atop\chi_{e}\right) \;\; ,
$$
\begin{equation}
\varphi_{e}=\left(0\atop
  g(z)\exp\left(-\rho^{2}/4a_{H}^{2}\right)\right) \;\; ,
\chi_{e}=\left(0\atop
  if(z)\exp\left(-\rho^{2}/4a_{H}^{2}\right)\right) \;\; .
\label{20}
\end{equation}
Dirac equations for functions $f(z)$ and $g(z)$ look like:
\begin{equation}
\begin{array}{c}
  g_{z}-(\varepsilon+m_{e}-\bar{V})f=0 \;\; ,\\
  f_{z}+(\varepsilon-m_{e}-\bar{V})g=0 \;\; ,
\end{array}
\label{21}
\end{equation}
where $g_z \equiv dg/dz$, $f_z \equiv df/dz$. They describe the
electron motion in the effective potential $\bar V(z)$:
\begin{eqnarray}
\bar V(z) & = & -\frac{Ze^2}{a_H^2}\left[1-e^{-\sqrt{6m_e^2}|z|} +
e^{-\sqrt{(2/\pi)e^3 B + 6m_e^2}|z|}\right]\times\nonumber\\&
\times &\int\limits_0^\infty
\frac{e^{-\rho^2/2a_H^2}}{\sqrt{\rho^2 + z^2}}\rho d \rho \;\; .
\label{22}
\end{eqnarray}
Intergrating (\ref{21}) numerically we find the dependence of $Z_{\rm
cr}$ on the magnetic field with the account of screening. The results
are shown in Fig. 2. For the given nucleus to become critical larger
magnetic fields are needed and the nuclei with $Z < 52$ do not become
critical.

\begin{figure}[h!]
  \centering
  \includegraphics[scale=0.4]{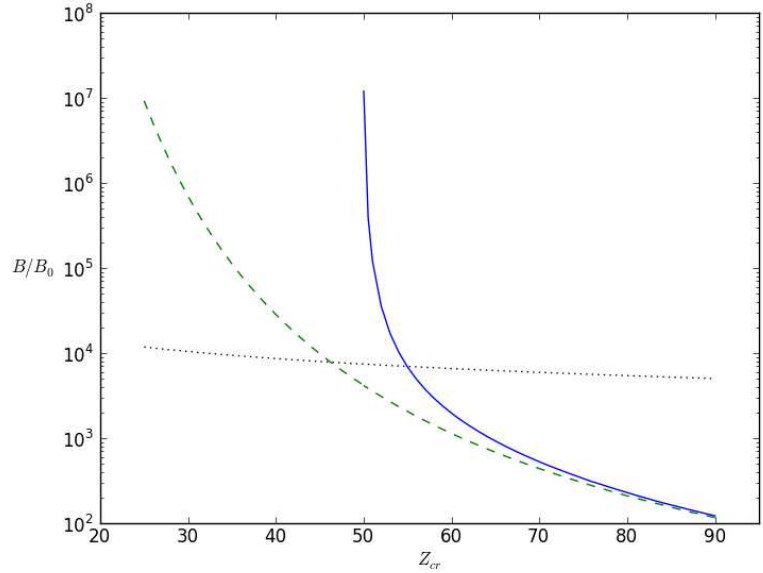}
  \caption{The values of $B^{Z}_{cr}$: a) without screening according
    to \cite{ORS}, dashed (green) line; b) numerical results
    with screening, solid (blue) line. The dotted (black) line corresponds
    to the field at which $a_{H}$ becomes smaller than the size of the
    nucleus.}
  \label{fig:Bcr}
\end{figure}

\section*{Acknowledgments}

I am grateful to the organizers for wonderful time in Nor-Amberd
and Tbilisi and to my coauthors Sergei Godunov and Bruno Machet
for helpful collaboration. I was partly supported by the grants
RFBR 11-02-00441, 12-02-00193, by the grant of the Russian
Federation government 11.G34.31.0047, and by the grant
NSh-3172.2012.2.


\begin{thebibliography}{0}
\bibitem{1}
M.I. Vysotsky, {\it JETP Lett.} {\bf 92}, 15(2010).
\bibitem{2}
B. Machet and M.I. Vysotsky, {\it Phys. Rev. D} {\bf 83}, 025022
(2011).
\bibitem{3}
S.I. Godunov, B. Machet, and M.I. Vysotsky, {\it Phys. Rev. D}
{\bf 85}, 044058 (2012).
\bibitem{4}
A.E. Shabad, V.V. Usov, {\it Phys. Rev. Lett.} {\bf 98}, 180403
(2009); {\it Phys. Rev. D} {\bf 77}, 025001 (2008).
\bibitem{VB}
V.B.Beresteckii, {\it Proceedings of LIYaF Winter School} {\bf 9,
part 3}, 95 (1974).
\bibitem{ZP}
Ya.B. Zeldovich, V.S. Popov, {\it UFN} {\bf 105}, 403 (1971);\\
W. Greiner, J. Reinhardt, {\it Quantum Electrodynamics} (Springer-Verlag, Berlin, Heidelberg, 1992);\\
W. Greiner, B. M\"{u}ller, and  J. Rafelski, {\it Quantum
Electrodynamics of Strong Fields} (Springer-Verlag, Berlin,
Heidelberg,1985).
\bibitem{ORS}
V.N. Oraevskii, A.I. Rez, and V.B. Semikoz, {\it Zh. Eksp. Teor.
Fiz.} {\bf 72}, 820 (1977) [{\it Sov. Phys. JETP} {\bf 45}, 428
(1977)].
\end{thebibliography}
\end{document}